\documentclass[letterpaper,twocolumn,10pt]{article}
\usepackage{usenix2019_v3}
\usepackage{hyperref}
\usepackage{tikz}
\usepackage{amsmath}
\usepackage{multirow}
\usepackage{array}
\usepackage{booktabs}
\usepackage{subfig}
\usepackage{adjustbox}
\usepackage{float}
\usepackage{caption}

\let\OLDthebibliography\thebibliography
\renewcommand\thebibliography[1]{
  \OLDthebibliography{#1}
  \setlength{\parskip}{0pt}
  \setlength{\itemsep}{0pt plus 0.3ex}
}
\begin{document}
%-------------------------------------------------------------------------------
%don't want date printed
%\date{}

\title{\Large \bf A First Step Towards Understanding Real-world Attacks on IoT Devices
}

%for single author (just remove % characters)
\author{
{\rm \quad Armin Ziaie Tabari} \qquad \quad  {\rm Xinming Ou} \\
aziaietabari@mail.usf.edu\quad xou@usf.edu \\
Department of Computer Science and Engineering\\
University of South Florida\\
Tampa, FL, USA
} % end author

\maketitle

\begin{abstract}

  With the rapid growth of Internet of Things (IoT) devices, it is
  imperative to proactively understand the real-world cybersecurity
  threats posed to them. This paper describes our initial efforts
  towards building a honeypot ecosystem as a means to gathering and
  analyzing real attack data against IoT devices.  A primary condition
  for a honeypot to yield useful insights is to let attackers believe
  they are real systems used by humans and organizations. IoT devices
  pose unique challenges in this respect, due to the large variety of
  device types and the physical-connectedness nature. We thus create a
  multi-phased approach in building a honeypot ecosystem, where
  researchers can gradually increase a low-interaction honeypot's
  sophistication in emulating an IoT device by observing real-world
  attackers' behaviors.  We deployed honeypots both on-premise and in
  the cloud, with associated analysis and vetting infrastructures to
  ensure these honeypots cannot be easily identified as such and
  appear to be real systems. In doing so we were able to attract
  increasingly sophisticated attack data. We present the design of
  this honeypot ecosystem and our observation on the attack data so
  far. Our data shows that real-world attackers are explicitly going
  after IoT devices, and some captured activities seem to involve
  direct human interaction (as opposed to scripted automatic
  activities). We also build a low-interaction honeypot for IoT
  cameras, called Honeycamera, that present to attackers seemingly
  real videos. This is our first step towards building a more
  comprehensive honeypot ecosystem that will allow researchers to gain
  concrete understanding of what attackers are going after on IoT
  devices, so as to more proactively protect them.

  % To identify these threats, it is better to first understand what those attackers want, and how they breach our network. This will finally help to have a safer and more secured environment. This paper, based on the honeypots that were studied for more than one year, introduces a novel honeypot to monitor the camera attacks.

\end{abstract}

%%% Local Variables:
%%% mode: latex
%%% TeX-master: "main"
%%% End:

\section{Introduction}
\label{sec:intro}

%\hspace{\parindent}
Over the past few years, a variety of devices used by people on a
daily basis has found their way to the Internet. IoT
has become one of the most hyped terminologies in
industry. According to IDC research~\cite{bigdata}, nowadays people
use various electronic devices at least three of which are usually
connected to the Internet. This number is estimated to increase to 10
devices per person in the near future. Gartner~\cite{gartner} expects
the world to see more than 20 billion IoT devices by 2020. Due to the
increase in IoT device usage, attacks on them have also
increased. For instance, more than 20\% of companies around the world
have experienced at least one IoT-related attack in the past few
years~\cite{cisco_2019,arubanetworks}.

  Historically cyber-attacks have mostly manifest as data breaches or
compromised devices used as spamming or DDoS agents.
% breaches are usually aimed at harming systems of
% vital companies in industry, personal computer devices, banks,
% automated vehicles and smartphones, to mention but a few.
% Also, there are many examples of serious and significant damages they have
% caused.  However, since nowadays, IoT devices play a key role in most
% people’s life, the widespread use of them has made cyber-attacks more
% dangerous. While the target of such attacks used to be a very specific
% group of people and companies in the past, many more people need to  worry
% and be aware of these possible threats.
The emerging of IoT devices could significantly change the landscape
of cyber-attacks both in terms of motives and methods. Due to the much
higher level of intimacy IoT devices possess to people's life, attacks
on them could result in much more devastating consequences compared
with cyber-attacks in the past.  Not only do they affect more people,
but the range of possible threats has also expanded. For instance, if
cyber criminals break into an IoT camera, they can invade people's
privacy at an unprecedented level.  Such attacks can even put people's
life in real danger (imagine an intruder takes control of an
autonomous vehicle).

Further exacerbating the situation is the repeated pattern in the IoT
industry where speed to market dominates concern for security.  Most
IoT devices have simple accessible vulnerabilities including default
username and password and open telnet/ssh port, to name just
two. Those devices are often installed in weak or unsecured networks
like home or public area.  We are unfortunately at a time when
exposure to attacks against IoT devices has become a reality, if not
worse compared to traditional computing systems.  According to
Symantec's report~\cite{symantec_2018}, the number of IoT attacks
dramatically increased in 2017. They identified 50,000 attacks which
had an increase of 600\% compared to 2016. New attacks such as
VPNFilter~\cite{vpnfilter}, Wicked~\cite{wicked},
UPnProxy~\cite{akamai_2018}, Hajime~\cite{edwards2016hajime},
Masuta~\cite{newsky_security_2018} and
Mirai~\cite{antonakakis2017understanding} botnet also show that
adversaries are continuously improving their skills to make these
forms of attacks even more sophisticated.  However, there is currently
very little systematic study on the nature and scope of those attacks
carried out in the wild.  So far, large-scale attacks on IoT devices
seen in the news have been mostly about using IoT in DDoS attacks
(e.g., the Mirai attack~\cite{antonakakis2017understanding}). It is
important to understand in fuller scope what activities attackers are
engaging with IoT devices and what their motives could be.

A honeypot is a device set up for the purpose of attracting cyber
attack activities. It is typically an Internet-facing device with
either emulated or real systems for attackers to target. Since these
devices are not meant to have other useful purposes, any access to
them are deemed malicious. Honeypots have helped security researchers
for a long time to understand various types of attacker behaviors.  By
analyzing data (network logs, downloaded files, etc.) captured by
honeypots, researchers can uncover new tools and methods used by
hackers, attack trends, and zero-day vulnerabilities. This information
is highly valuable to improve cyber security measures, especially when
organizations are resource-strapped when it comes to fixing security
vulnerabilities.

In this paper we present our first step towards a comprehensive
experimentation and engineering framework for capturing and analyzing
real-world cyber-attacks on IoT devices using honeypots. There are two
main challenges for creating IoT honeypots that can yield useful
data for research.

\begin{enumerate}
\item The types of different IoT devices are vast, each of which
  has unique features that an attacker may wish to access. It is
  infeasible to build one honeypot system that can capture even a significant
  portion of all IoT devices. Thus, we adopt a {\it multi-faceted} approach
  to IoT honeypot engineering. We both adapt existing off-the-shelf
  honeypot systems and build ones from scratch to create a variety of
  honeypot systems for attackers to target.

\item The specific nature of attackers' activities towards IoT devices
  is largely unknown at this point, and there could be very different
  focuses on the attacker's side. Moreover, the richness of response
  from an IoT device is much greater than traditional IT systems due
  to the interaction with the physical world. For example, an IoT
  camera will need to return some real video to look like a real device.
  It would require significant amount of engineering work to emulate
  those different types of responses for different devices. Thus, we adopt
  a {\it multi-phased} approach where the sophistication of the emulated
  responses are gradually increased as data is gathered and analyzed
  to understand what the attackers might be going after.

\end{enumerate}

In the remainder of the paper, we describe this multi-faceted,
multi-phased approach in building a {\it IoT honeypot ecosystem},
where various types of honeypots, deployed both on-premise and in the
cloud, work in concert with a vetting system (to ensure the honeypot
device looks like a real device) and data analytics infrastructure
(for collecting and analyzing the captured data). Our ultimate goal is
to attract real human attackers and understand their motives and {\it
  modus operandi}, so as to inform research on securing IoT devices in
the real world.

%\begin{comment}
%In this work, different honeypots were implemented to mimic IoT
%devices, and finally, a state-of-the-art honeypot for camera devices
%was designed. This paper is structured as follows: Related work is
%discuss in section 2. Experimental setup is described in section 3. It
%is followed by the findings of this study in section 4, and finally
%section 5 brings the conclusion

%\end{comment}

%%% Local Variables:
%%% mode: latex
%%% TeX-master: "main"
%%% End:

\section{Honeypot Background and Related Work}
\label{sec:related}

% Creating an effective cyber-security procedure or product needs a
% thorough understanding of the existing and possible threats. Internet
% of Things networks have become a new interesting target for
% adversaries to perform their malicious activities. It is highly
% crucial to understand what those attackers want from these networks,
% and why they changed their attention from standard systems to this new
% area.
The first honeypot was introduced in 2000~\cite{honsurv}.  Honeypots
can be categorized into two classes: Low-interaction honeypot and
high-interaction honeypot.  Low-interaction honeypots only emulate some
services such as SSH or HTTP, whereas high-interaction honeypots
provide a real operating system with lots of vulnerable
services~\cite{honsurv}.
% For more detailed information, please refer to
Honeypots are also categorized based on their
purpose~\cite{spitzner_2001}.  Production honeypots help companies
mitigate possible risks, and research honeypots provide new
information for the research community.  Alba et
al.~\cite{alaba_othman_hashem_alotaibi_2017} conducted a survey of
existing threats and vulnerabilities on IoT devices. The first time
IoT devices were used as a platform for large Internet-scale attack
dates back to the summer of 2016, when the French hosting company OVH
was targeted with the first wave of Mirai
attacks~\cite{antonakakis2017understanding}. In the follow-up attack
in October 2016, Mirai brought down the Dyn DNS provider which at the
time was hosting major companies’ websites including Twitter, Github,
Paypal and so on.

Luo et al.~\cite{luo2017iotcandyjar} designed an
intelligent-interaction honeypot for IoT devices called IoTCandyJar.
It actively scans other IoT devices around the world and sends some
part of the received attacks to these devices. This approach provides
an easy way to have a realistic interaction with adversaries,
although it may inadvertently
bring harm to otherwise innocent IoT devices exposed on the
Internet. Guarnizo et al.~\cite{guarnizo2017siphon} proposed a
high-interaction IoT honeypot. They used real IoT devices to capture
and analyze malicious activities. Another solution proposed by
Pa et~al.~\cite{minn2015iotpot} %\oxm{check last name}
was a combination of low interaction
honeypots with sandbox-based high-interaction honeypots %\oxm{high-interaction?}
implemented as IoTPot. IoTPot
only monitors telnet-based attacks. A honeypot for monitoring Zigbee
protocol was offered by Dowling et~al.~\cite{dowling2017zigbee}. Logs
from a Zigbee honeypot were analyzed in this work. Six types
of honeypots were deployed by Chamotra
et~al.~\cite{chamotra2016honeypot} to understand IoT attacks in the
broadband network.
Compared to the prior work mentioned above, our main contribution
is the design, implementation, and deployment of a honeypot ecosystem that addresses
the challenges of capturing useful attack data on IoT devices (section~\ref{sec:intro}).
%\oxm{Need to think more about the differentiating factors}
% we designed an ecosystem for honeypots which has a multi-phased process. During
% this process, a low-interaction honeypot will become more sophisticated which
% can attract more advanced attacks to itself. Moreover, a low-interaction camera
% honeypot was introduced, and we observed how adversaries attack this honeypot.

%%% Local Variables:
%%% mode: latex
%%% TeX-master: "main"
%%% End:

\section{A Honeypot Ecosystem}
\label{sec:eco}

\begin{figure*}[]\centering
  \includegraphics[width=\linewidth]{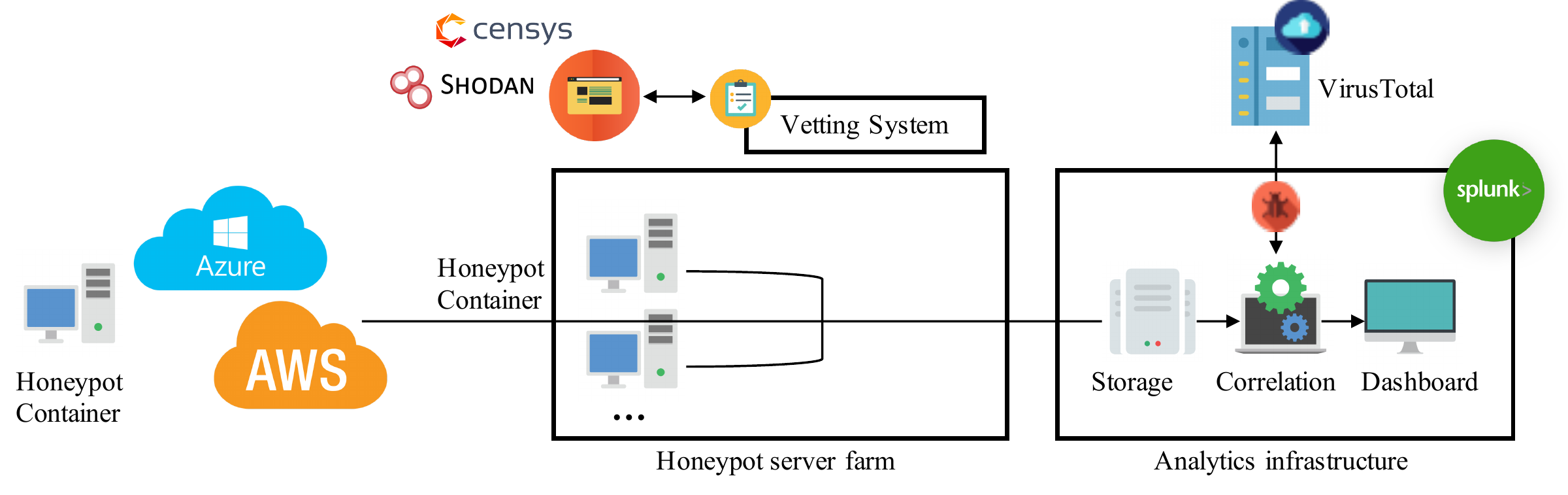}
  \caption{First Step Towards a Honeypot Ecosystem}
  \label{fig:Bigpicture}
\end{figure*}

%\hspace{\parindent}
For honeypots to be useful for IoT security
research, it is not sufficient to just have boxes running the various
emulated or real IoT systems. These boxes need to be organized in a
way that allow intelligent adaptation on the way they respond to
different types of traffic, so an attacker can be ``hooked'' and made
interested in further exploring it. The deeper an attacker becomes
interested in a device, the more sophisticated it needs to be in
``fooling the attacker'' into thinking it is a real device. This
inevitably becomes an arms race, the effectiveness of which lies in
the return of useful insights per the amount of engineering efforts
spent. Our goal is to create a carefully designed ecosystem where a
variety of honeypot devices working together with a vetting and
analysis infrastructure, so that we can achieve good ``return on
investment.''

Figure~\ref{fig:Bigpicture} illustrates our first step towards building
such an ecosystem. There are three distinct component:
1) honeypot server farms (on premise and in cloud),
2) a vetting system to ensure it is adequtely difficult for an adversary to
detect the honeypot device is a honeypot, and
3) an analysis infrastructure that is used to collect and analyze the captured data.

\subsection{Honeypot Server Farms}
\label{sec:serverfarms}

\begin{figure}[htb]\centering
  \includegraphics[width=.8\linewidth]{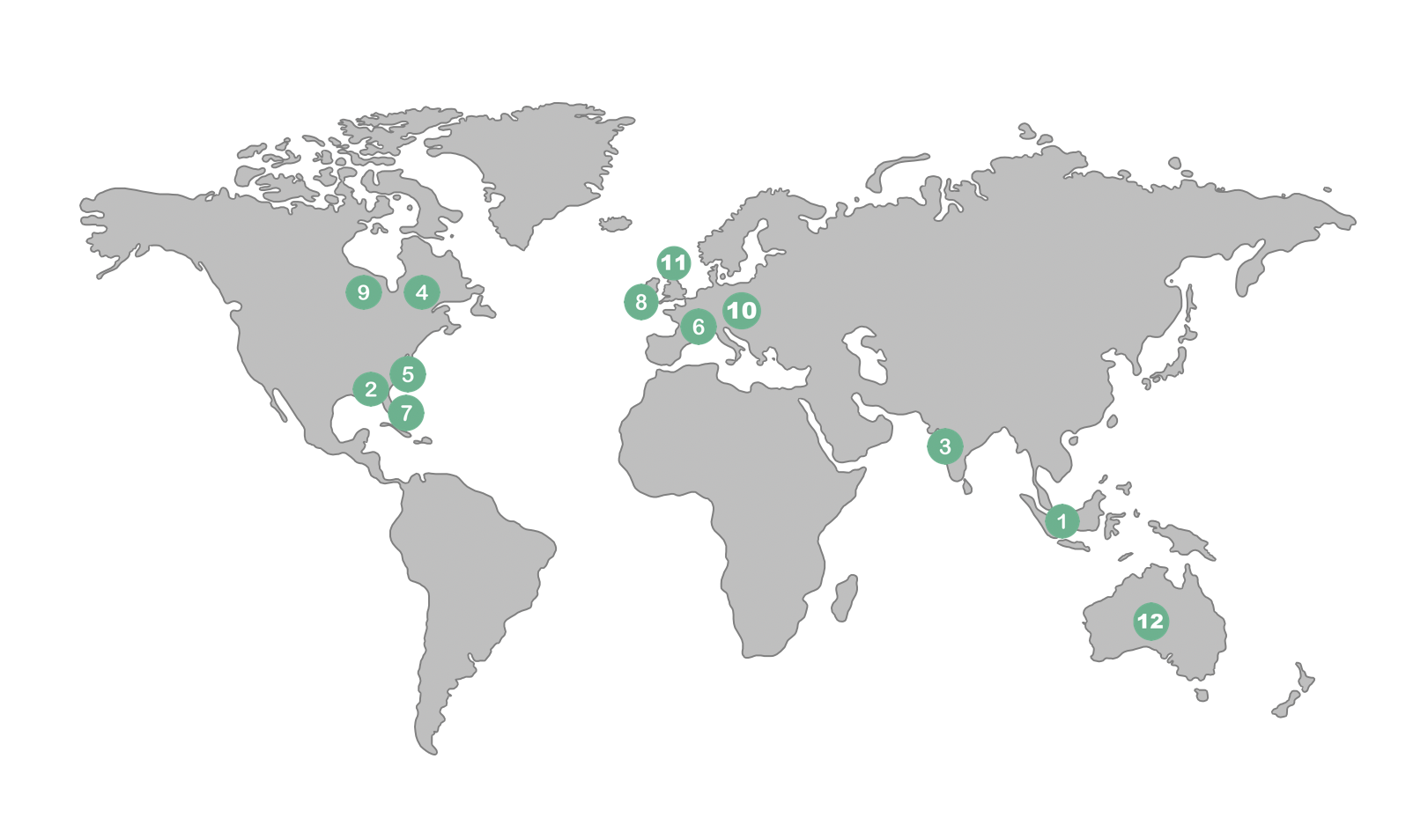}
  \caption{Honeypot Deployment Locations}
  \label{fig:locationHPs}
\end{figure}

The honeypot server farms host the honeypot devices. We use both
on-premise server and cloud instances from AWS~\cite{amazon} and
Azure~\cite{azure} in multiple countries, to create a wide geological
coverage. Figure~\ref{fig:locationHPs} shows the geographical
locations of the deployed honeypot instances in our server
farms. The countries include France, United Kingdom, India, Australia,
Canada and Singapore. The on-premise server farm consists of a
PowerEdge R830 server with 256 GB RAM running as VMware ESXi server,
and a Synology NAS server for storing logs. Three Fedora instances and
one Windows instance run on the ESXi server. The Windows instance and
two of the Fedora instances are used for deploying honeypots. The
third Fedora instance is used to run Splunk~\cite{splunk} to support
data analytics.  The instances in AWS and Azure run Ubuntu or Windows
depending on the type of honeypots deployed.  At this stage of our
research, we only use low-interaction honeypots. For Fedora and Ubuntu
instances we use docker containers to run honeypots in them.  The
generated logs are sent through syslog protocol to splunk. Networking
controls are configured through security groups so that ports open for
communication between entities within the honeypot ecosystem are not
visible by Internet attackers.

As explained in Section~\ref{sec:intro}, given that different IoT
devices have different specifications and configurations, each
honeypot needs to be designed and configured in a unique way.  We
adopt a multi-faceted approach to building the various honyepot
instances.  We both use off-the-shelf honeypot emulators and adapt
them, and build specific emulators from scratch.

\subsubsection{Off-the-shelf Honeypots}
\label{sec:ots}

Many popular off-the-shelf honeypots emulate general services and
protocols that are not specific to IoT. However since many IoT devices
have those services, it is still useful to adapt these existing
honeypots for studying IoT attacks.  We evaluated various open-source
and commercial honeypots and selected three off-the-shelf software to
use in the first step: {\it Cowrie}~\cite{cowrie}, {\it
  Dionaea}~\cite{dionaea} and {\it KFSensor}~\cite{kfsensor}.
In the rest of this section,
a brief introuduction of these honeypots is presented.

\paragraph{Cowrie} is a low-interaction honeypot\footnote{The Cowrie
  author uses the term ``medium interaction'' honeypot; but it falls
  within the low interaction category based on the definition introduced
  in Section~\ref{sec:related}.} that attempts to
imitate SSH and telnet services to attract adversaries and capture
their interaction. Cowrie provides a fake file system, a fake ssh shell
and is also able to capture files from input. It can log
every activity in JSON format for ease of
analysis~\cite{cowrie}. Since many IoT devices still use telnet
and SSH protocols for management purposes, Cowrie is a
useful honeypot for understanding some facets of attacks on IoT devices.
We run Cowrie on Debian inside a docker container.

\paragraph{Dionaea} is a low-interaction honeypot that emulates
various vulnerable protocols commonly found in a Windows system.
The main goal of this honeypot is to
lure adversaries and capture their malicious files such as worms and
malware. Dionaea is able to simulate various protocols including
HTTP, MYSQL, SMB, MSSQL, FTP, and MQTT. This honeypot also
logs all the detected events in JSON format or inside a SQLite
database. Dionaea was released in 2013 and is a useful tool
for traping malware that exploits vulnerabilities~\cite{dionaea}.
We run Dionaea on Debian inside a docker container.

\paragraph{KFSensor} is a commercial Intrusion Detection System (IDS)
that acts as a low-interaction honeypot to attract potential
adversaries and record their activities. It is a windows-based
honeypot.  By acting as a bait, KFSensor draws the attention of
adversaries from the real systems to itself and provides valuable
intelligence for both research and operation.  KFSensor also has some
features useful for this work: managing the sytem remotely, easy
integration with other IDSs like Snort~\cite{snort}, and emulating
Windows network protocols~\cite{kfsensor}.  Since Windows has
substantial footprint as an IoT operating systems, both Dionaea and
KFSensor can shed light on attacks on IoT systems. We run KFSensor
in the Windows VMs on our server farms.

\subsubsection{HoneyCamera}
\label{sec:hc}

\begin{figure}[h]
  \includegraphics[width=\linewidth]{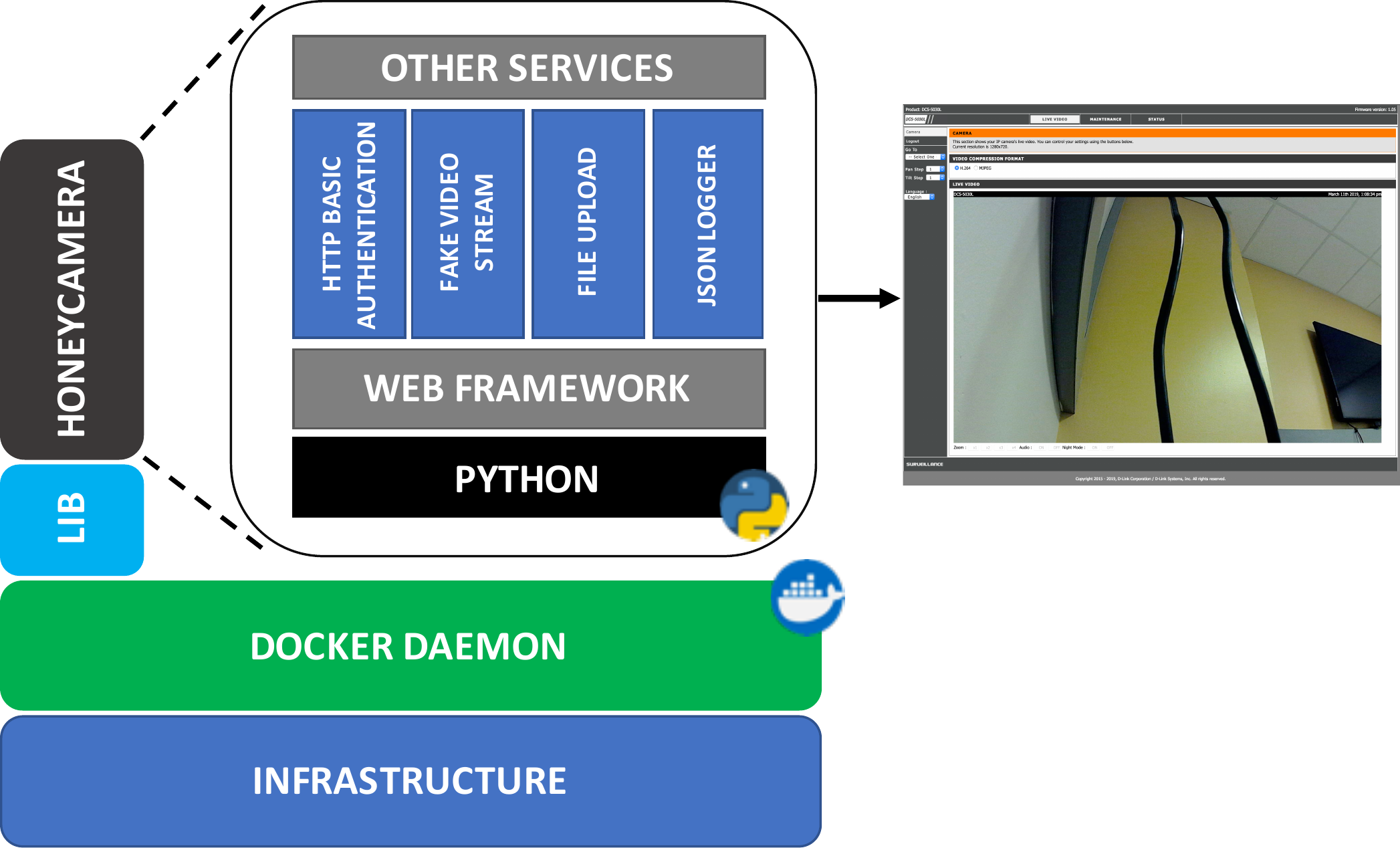}
  \caption{HoneyCamera architecture}
  \label{fig:honeycamera}
\end{figure}

To capture attacks on specific IoT devices, we build a honeypot for
IoT camera and coined it {\it HoneyCamera}.  This is our first step
towards understanding what attackers are going after for various types
of specific IoT devices. Figure~\ref{fig:honeycamera} shows the
architecture of this honeypot. Honeycamera is a low-interaction
honeypot for D-Link IoT cameras. We implemented it using python3. To
achieve this objective, we studied a D-Link camera and its responses
to various types of inputs were captured carefully. Honeycamera uses
basic authentication for login and repeatly plays a few seconds' of
real video as a fake video stream from the emulated camera device.  We
also built six different pages which emulate the various services
supported by this IoT camera, such as password changing, reading
network information, and creating new users.  This helps us understand
the behavior of the adversaries. We also created a fake firmware
upload service to capture and analyze tools and exploits used during
attack sessions. Honeycamera logs all activities in JSON format.
HoneyCamera runs in Clear Linux~\cite{clearlinux} that in turn runs
inside a docker container.

% One of the technologies that has recently gained popularity, approval and support in the software development field is containerization technology. When a software is moved among different development environments, it is possible for it to lose functionality. Containers are used to ensure that this problem will not happen. Considering advantages of both low interaction honeypots and the containerization architecture, they can work together to make an effective tool which is capable of detecting attacks as soon as possible. So, it was decided to use this architecture along with honeypots. Each honeypot's container includes all of the dependencies, configuration files and libraries needed to run bug-free and without error. For this purpose, \textbf{Docker}, one of the most best-known containerization ecosystems, was utilized. Docker supports different types of architecture which was a good choice for the installed honey-farm. Also, its containers use less memory and CPU which made Docker perfect for this project.

\subsection{Honeypot Vetting}
\label{sec:vetting}

A honeypot is valuable only as long as it remains undetectable, i.e.,
unknown to the attacker as a fake system.  This is inherently a hard
task since honeypots (especially low-interaction ones) will inevitably
fail to demonstrate some observable features only a real system can
possess, or present ones a real system will never show.
% At a minimum, we need to make sure publicly available fingerprinting
% systems cannot identify our honeypots as such.
% Being publicly available, having well-known
% characteristics and existing fingerprinting techniques to identify
% honeypots make deploying them a very difficult task.
% For the purpose of
% having a hard-detectable honeypot, an effective vetting procedure is
% necessary.
The goal of the vetting process is to identify any information leakage
which leads to the identification of a honeypot and provide remedies
accordingly. We use the server farms in the cloud as a testbed to
try different fingerprinting approaches to first make sure our
honeypots cannot be easily detected.
We use some manual
and automatic fingerprinting method (e.g., through Metasploit~\cite{rapid7}).
We use Shodan~\cite{shodan}, an IoT search engine
that can be used to search for IoT devices on the Internet.
Shodan provides information such as service banners and metadata, and
a {\it honeyscore} in the range from 0 to 1 (1 indicates honeypot while 0 means
real system). This score provides a preliminary insight into how good the honeypot
impersonates a real device. We use Censys~\cite{censys}, another
IoT search engine, to help analyze our honeypot instances to make sure
they look like the real ones they imitate. Finally, and
most importantly, from the captured data inside honeypots, attackers'
fingerprinting approaches can be identified.  Based on this insight we
can create solutions to make them ineffective. This is
built in our multi-phased experimentation process which will be
explained in further details in Section~\ref{sec:multiphased}.

\subsection{Data Analytics Infrastructure}
The success of a honeypot depends on two factors: 1) the way the
honeypot software is developed and implemented; and 2) the log
analysis process. Carefully analyzing
the logs is as important as the honeypot development and
implementation. We use Splunk~\cite{splunk} for log management and analysis.
Splunk supports composing various queries using its domain-specific language
that can be leveraged to achieve various analysis purposes in this work.
All the logs captured from our honeypots are sent to the Splunk
server for further analysis.
We created a Splunk app to extract valuable information from the collected logs.
Some example analyses done by the app are
identifying the combinations of username and password used by
attackers, analyzing locations of the attacks, detecting the most
and least command executed during attack sessions, analyzing
downloaded files and sending them directly to VirusTotal~\cite{vt},
storing the results and checking attackers’ IPs through DShield~\cite{dshield}
and AbuseIPDB~\cite{abuseipdb}, and so on.
% These are only some of the most
% important features that were put in this log management application.
% has a real-time data collection and visualization
% functionality as well as streamline investigations, dynamically search
% and analyzing logs and leverage of having artificial intelligence and
% machine learning embedded in it. Considering such a wide range of
% operations, Splunk was a right option for managing honeypots'
% logs.

%%% Local Variables:
%%% mode: latex
%%% TeX-master: "main"
%%% End:

\section{Multi-phased Deployment/Experimentation}
\label{sec:multiphased}

% For the purpose of this project, a multi-faceted approach was chosen.  Although, these honeypots provided satisfactory results, to have more specific information on IoT attacks, {\it HoneyCamera} designed and deployed along with others. Each one of these honeypots emulated at least one IoT device’s protocol, including {\it HTTP} and {\it SSH}, to name just two.
We employ a multi-phased approach, as explained in
Section~\ref{sec:intro}, to introduce sophistication in how our
honeypots respond to attackers' traffic, based on traffic collected
previously.  In the first phase, we simply deploy the honeypot at hand
and receive attack traffic.
% whatever at our disposal had the least changes from the original software to start
% receiving attacks.
From this point, the honeypot ecosystem starts to collect data, and
the analyzed data inform the creation of the subsequent phases by
understanding what cybercriminals are looking for, so we can provide
emulated responses accordingly. This second phase goes through many
iterations until we are satisfied with the insights we gained and the
attackers' behaviors elicited. Then in the third phase these insights
are used to create more advanced low-interaction honeypots. We present
this multi-phased process from three facets that our honeypots attempt
to capture about IoT attacks: attacks through login service to obtain
a command shell, windows service attacks resulting in malware download,
and IoT camera attacks.

\begin{description}
\item[HoneyShell] We use the Cowrie honeypots to emulate vulnerable
IoT devices with open SSH (port 22) or telnet (port 23). Cowrie can be
configured to emulate different types of OS
platforms. Busybox~\cite{busybox} is a Linux distribution popular
among IoT devices. Thus we configure our Cowrie honeypots to emulate
Busybox.  We created three Cowrie honeypots for the three
phases.
\begin{itemize}
\item {\it Phase 1}: the system is allowed to accept every possible
combination of usernames and passwords. Two of these honeypots were
deployed, one in Singapore (cloud) and the other on-premise.

\item {\it Phase 2}: deployed on-premise and started after 6
months. The top 30 username/password combinations with at least one
command execution after login in phase~1 were selected.  All other
username/password combinations would fail in this version of
honeypot. Moreover, the emulation mechanisms are configured so that
those commands will produce a meaningful response to attackers, e.g.,
new usernames and file systems were added in the configuration.  Also,
phase 1 logs were analyzed to identify possible fingerprinting
techniques by attackers so they will be taken care of in phase 2.
Examples include \textit{file} command's response added to the
honeypot configuration.

\item {\it Phase 3}: all information collected before is used to
create a more sophisticated honeypot. As a result, a complex password
was generated, and the number of possible successful login
combinations was limited to only one. Since the password was set to be
complicated, every successful login would imply that it was possibly
done by a real hacker which consequently could provide valuable
information.
\end{itemize}

\item[HoneyWindowsBox] We use Dionaea to emulate IoT devices based
  on Windows platform. These attacks mostly result in malware
  being downloaded on the device. It would involve non-trivial
  work to further emulate the downloaded malware's behavior inside
  a honeypot; we leave this for future work. In this work, we use
  phase 2 of this honeypot to apply our vetting system to ensure
  they are not easily identifiable as honeypots.

\begin{itemize}
\item {\it Phase 1}: a default version of Dionaea was deployed in
the cloud. Our vetting system quickly identified it as a honeypot.

\item {\it Phase 2}: the various services were broken down into
  two different combinations. The first honeypot
  has FTP, HTTP and HTTPS while the second only contains SMB and MSSQL services.
  We deployed the two versions to three locations (India, Canada, and on-premise).
  The IPs popped up as a real system in Shodan. We also deployed KFSensor
  during this step. Paris and on-premise were chosen as the locations
  and all deployed instances passed our vetting system.
\end{itemize}

\item[HoneyCamera] % is the last implemented honeypot. Different cameras
  % have different available ports.
  This is our latest effort in emulating behaviors of a more specific IoT device.
  We plan to use data collected in phase 1 to inform the HoneyShell configuration
  in phase 2. This phase is still work in progress.
  
\begin{itemize}
\item {\it Phase 1}: three honeypots were deployed. Two in Sydney and
Paris had only port 8080 open while the other in London had port
80. The first two honeypots were used to emulate D-Link DCS-5020L and
the other one to imitate D-Link DCS-5030L camera.

\item {\it Phase 2}: a combination of HoneyCamera and HoneyShell will
  enable us to elicit attacker's behaviors that involve both common
  Unix commands and camera-specific commands.
\end{itemize}
\end{description}

%%% Local Variables:
%%% mode: latex
%%% TeX-master: "main"
%%% End:

\section{Experimentation and Data Analysis}
\label{sec:dataanalysis}

%\hspace{\parindent}
A total number of 18,972,475 hits were captured by our
honeypot ecosystem over a period of one year. As shown
in table~\ref{table:1}, HoneyShell attracted the most hits.
This information is described in detail in the rest of this section.

\begin{table}[htb]\centering
\caption{Number of Hits based on Different Honeypots}
\begin{tabular*}{\columnwidth}{@{\extracolsep{\fill}}ccc}
\toprule
\bfseries Honeypot & \bfseries Up Time & \bfseries \# of Hits\\
\midrule
HoneyShell       & 12 months  & 17,343,412 \\ \midrule
HoneyWindowsBox    & 7 months   & 1,618,906  \\ \midrule
HoneyCamera    & 1 months          & 10,157      \\
\bottomrule
\end{tabular*}

\label{table:1}
\end{table}

%-------------------------------------------------------------------------------
\subsection{HoneyShell}
\begin{figure}[htb]\centering
  \includegraphics[width=.8\linewidth]{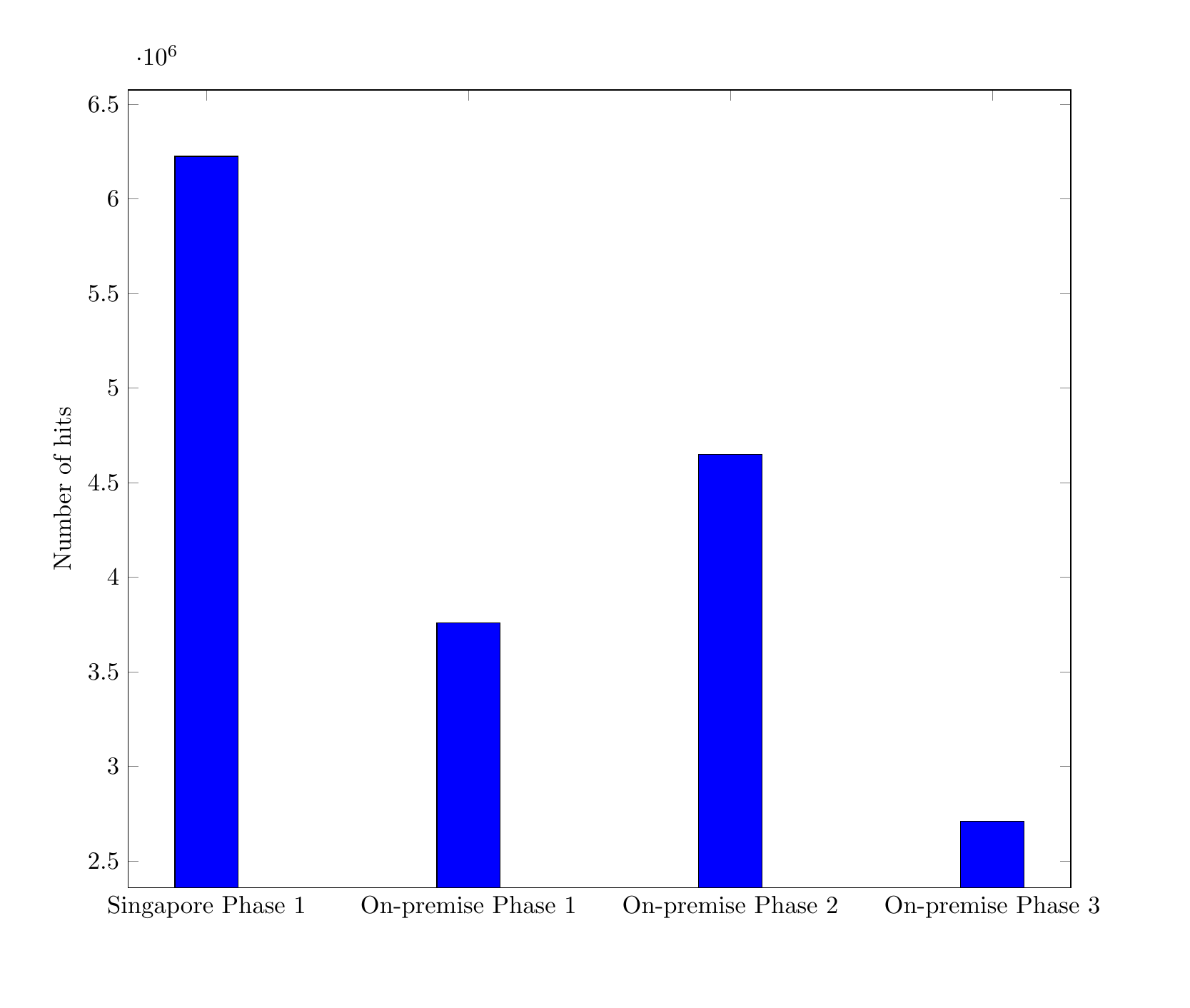}
  \caption{Cowrie Honeypot Hits per Location/Phase}
  \label{fig:cowrie}
\end{figure}

\begin{figure}[t!]\centering
  \includegraphics[width=.8\linewidth]{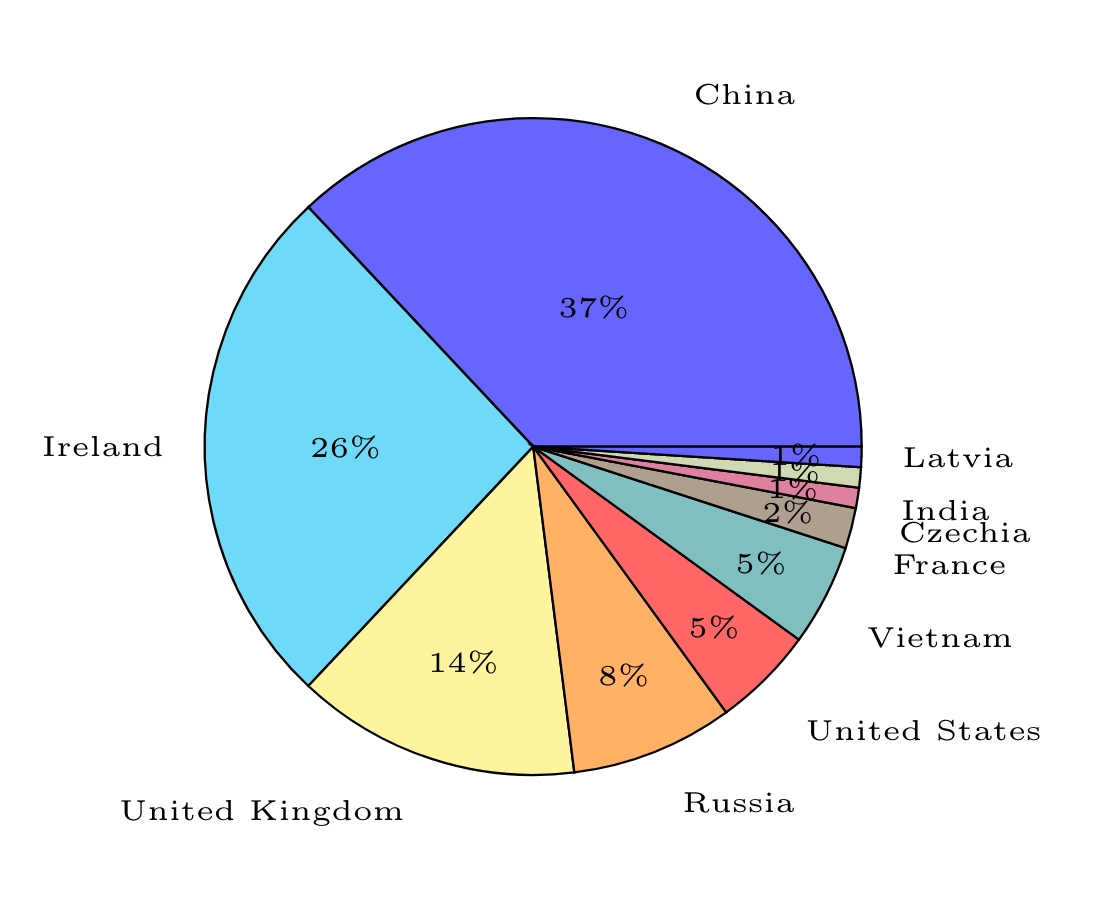}
  \caption{Top 10 Countries with the Most Connections}
  \label{fig:pie}
\end{figure}

%-------------------------------------------------------------------------------
%\hspace{\parindent}
Cowrie honeypots were able to capture the largest portion of the hits
during this period. Figure~\ref{fig:cowrie} represents the number of
hits based on locations and phases. It is notable that the on-premise
phase 2 honeypot captured more hits in 6 months' time than the on-premise
phase 1 honeypot did in a year, clearly showing the effectiveness of the multi-phased
approach. Figure~\ref{fig:pie} shows that
the majority of connections came from China, Ireland
and the United Kingdom.


Furthermore, statistics shows that 15\% of the total number of hits
belong to successful logins. Most of these logins used random
combinations of username and password which shows that automated
scripts were used to find the correct authentications blindly. Table
\ref{table:2} represents the top 10 username/password combinations
that were used by attackers. % This table clearly shows that the
% combination of \textbf{\textit{admin}}/\textbf{\textit{1234}} was used
% the most by attackers. By a large margin,
% \textbf{\textit{root}}/\textbf{\textit{(empty)}} placed second. The
% combination of \textbf{\textit{admin}}/\textbf{\textit{(empty)}} also
% found its way to the table in third place.
The information seems to indicate that
attackers commonly look for high-value user
with a weak password. However, by looking into the database, some
other combinations such as
\textbf{\textit{university}}/\textbf{\textit{<florida>}},
\textbf{\textit{root}}/\textbf{\textit{university}} and
\textbf{\textit{university}}/\textbf{\textit{student}} were found
inside the on-premise honeypot (inside a university)
which indicates that attackers were aware of the organization's nature.
and tried to customize their attacks based on that.

\begin{table}[htb]\centering
\caption{Top 10 Username and Password combinations}
\begin{tabular*}{\columnwidth}{@{\extracolsep{\fill}}cccc}
%\begin{tabular}{*2c}
\toprule
{} & \bfseries Username/Password & \bfseries Occurrences\\
\midrule
{} & admin / 1234  & 975729 \\ \midrule
{} & root / (empty)        & 167869    \\ \midrule
{} & admin / (empty)  & 82018     \\ \midrule
{} & 0 / (empty)      & 62140     \\ \midrule
{} & (empty) / root   & 52780     \\ \midrule
{} & 1234 / 1234      & 50305     \\ \midrule
{} & admin / admin    & 39349     \\ \midrule
{} & admin / 1234567890    & 12444     \\ \midrule
{} & root / admin     & 10359     \\
\bottomrule
\end{tabular*}
\label{table:2}
\end{table}

In addition, only 314,112 (13\%) unique sessions were detected with at
least one successful command execution inside the honeypots. This
result indicates that only a small portion of the attacks executed
their next step, and the rest (87\%) solely tried to find the correct
username/password combination. A total number of 236 unique files
were downloaded into honeypots. 46\% of the downloaded files belong to
three honeypots inside the university, and the other
54\% were found in the honeypot in Singapore. Table~\ref{table:3}
demonstrates categorization of the captured malicious files by
Cowrie. VirusTotal flagged all these files as
malicious. DoS/DDoS executables were the most downloaded ones inside
honeypots. Attackers tried to use these honeypots as a part of their
botnets. IRCBot/Mirai and SHelldownloader were the second most
downloaded files. It shows that Mirai, which was first introduced in
2016, is still an active botnet and has been trying to add more
devices to itself ever since. Shelldownloader tried to download
various format of files that can be run in different operating
systems' architectures like \textit{x86, arm, i686} and
\textit{mips}. It should be highlighted that since adversaries were
trying to gain access in their first attempt, they would run all the
executable files. SSH scanner, mass scan and DNS Poisoning are
categorized in other sections of table \ref{table:4}.

\begin{table}[htb]\centering
\caption{Categorization of downloaded files}
\begin{tabular*}{\columnwidth}{@{\extracolsep{\fill}}cccc}
\toprule
{} & \bfseries Malicious Files Campaign & \bfseries Amount\\
\midrule
{} & Dos/DDos &59 \\ \midrule
{} & IRCBot/Mirai &40 \\ \midrule
{} & SHELLDownloader&40 \\ \midrule
{} & BACKDOOR & 36 \\ \midrule
{} & CoinMiner& 31 \\ \midrule
{} & Others& 30 \\
\bottomrule
\end{tabular*}
\label{table:3}
\end{table}

Besides downloading files, attackers tried to run different commands.
Table~\ref{table:4} shows the top 10 commands executed with their occurrence
number. By examining SSH version banner information, the difference
between human attacks and bot attacks was identified. While 61\% of
the attacks are carried by a bot, the rest (39\%) are likely human attacks.

\begin{table}[htb]
\centering
\caption{Top 10 Commands Executed}
\begin{tabular*}{\columnwidth}{@{\extracolsep{\fill}}cc}
\toprule
\bfseries Command & \bfseries Occurences\\
\midrule
cat /proc/cpuinfo   &  15453\\ \midrule
free -m   &  11344\\ \midrule
ps -x   &  11204\\ \midrule
uname -a   &  5965\\ \midrule
export HISTFILE= /dev/null  &  5949\\ \midrule
grep name &  3798\\ \midrule
/bin/busybox cp; /gisdfoewrsfdf &    1141\\ \midrule
/ip cloud print &   883\\ \midrule
\multicolumn{1}{c}{lspci | grep VGA | head -n 2 | tail -1 |} & \multirow{2}{*}{532} \\
\multicolumn{1}{c}{ awk \'\{print \$5\}\'} \\
\bottomrule
\end{tabular*}
\label{table:4}
\end{table}
%-------------------------------------------------------------------------------
\subsection{HoneyWindowsBox}
%-------------------------------------------------------------------------------

%\hspace{\parindent}
Dionaea was representing a vulnerable Windows operating system. Most
of the connections came from the United States followed by China and
Brazil. During the usage of Dionaea, 43 unique files were
captured. \textit{HTTP} was the protocol used the most by attackers.
\textit{FTP} and \textit{smb}
were also used to download malicious files. In addition, a noticeable
amount of \textit{SIP} communication was found in the process of
examination. SIP is mostly used by VoIP technology, and like other
services, it suffers from common vulnerabilities such as buffer
overflow and code injection. Collected data from
these honeypots was used to create a more realistic file system for
other honeypots.

KFSensor is an IDS-based honeypot. It listens to all ports and tries
to create a proper response for each request it receives. The
information gathered from this honeypot was also used to create a better
environment and file system for Dionaea.
%-------------------------------------------------------------------------------
\subsection{HoneyCamera}
%-------------------------------------------------------------------------------
\begin{table}[!t] \centering
\caption{Attack Types Executed inside HoneyCamera}
\begin{tabular*}{\columnwidth}{@{\extracolsep{\fill}} ccc} \toprule {}
& \bfseries Attack Type & {}\\ \midrule {} & {[}CVE-2013-1599{]} DLINK
Camera & {} \\ \midrule {} & Hikvision IP Camera - Bypass
Authentication & {} \\ \midrule {} & Netwave IP Camera - Password
Disclosure & {} \\ \midrule {} & AIVI Tech Camera - command injection
& {} \\ \midrule {} & IP Camera - Shellshock & {} \\ \midrule {} &
Foscam IP Camera - Bypass Authentication & {} \\ \midrule {} &
Malicious Activity & {} \\ \bottomrule
\end{tabular*}
\label{table:5}
\end{table}
\begin{figure}[!t]\centering
\includegraphics[width=.7\linewidth]{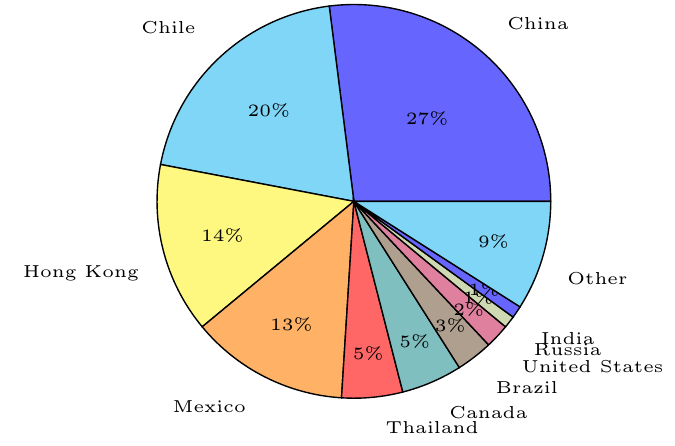}
  \caption{Top 10 countries with most attacks}
  \label{fig:honeycam}
\end{figure}
%\hspace{\parindent}
HoneyCamera was employed to emulate three IoT camera devices. As shown
in Figure~\ref{fig:honeycam}, most attacks captured inside HoneyCamera
were from China. Two types of malicious files were downloaded with
this honeypot: CoinMiner and Mirai. Analyzing the captured logs
reveals that this honeypot attracted many attacks specifically targeted
at IoT cameras. Below are some examples:

\begin{itemize}
  \item The first attack found was camera credential brute-force
(\textit{/?action=stream/snapshot.cgi?user=[USERNAME]\&\\pwd=[PASSWORD]\&count=0}). On
this attack, adversaries tried to find a correct combination of
username and password to get access to the video streaming service.
  \item The second attack found was trying to exploit \textbf{CVE-2018-9995}
vulnerability. This vulnerability allows attackers to bypass
credential via a \textbf{``Cookie: uid=admin''} header and get access
to the camera (\textit{/device.rsp?opt=user\&cmd=list}).
  \item A list of more attacks can be found in Table~\ref{table:5}.
\textit{D-Link}, \textit{Foscam}, \textit{Hikvision}, \textit{Netwave}
and \textit{AIVI} were only some of the targeted cameras inside this
honeypot.
\end{itemize}

In addition, attackers mostly (74\%) used \textbf{\textit{GET}}
protocol to communicate with the honeypot and 23\% used
\textbf{\textit{POST}} method.

%%% Local Variables:
%%% mode: latex
%%% TeX-master: "main"
%%% End:

\section{Conclusion}

In this paper we present  a multi-faceted and multi-phased approach in building
a honeypot ecosystem. Furthermore, a new low-interaction honeypot for camera
devices was introduced. Although this honeypot is still
wrok in process, it has provided much
insightful information. Analysis on the information captured during this work
shows that adversaries generally look for vulnerable IoT devices to exploit
them. Also, results indicate that in the same network , a more realistic and
well-configured low-interaction honeypot can attract more attacks
compared to honeypots which is configured poorly. Moreover, analysis of
HoneyCamera's logs shows that IoT camera devices have become an interesting
target for attackers. Different types of vulnerabilities were found in this
process.

%That is why the implemented honeypot in phase 2 captured more malware files compared to that of phase one. This result also shows that a well-known combination of username and password makes the target more valuable for adversaries, so they will try to execute their next step in the attacks.
%D-Link, Foscam, Hikvision, Netwave and AIVI were only some of the targeted cameras inside this honeypot.
%Using Cowrie honeypot to emulate an operating system for HoneyCamera will be the
%next step to capture more information from this honeypot. However, more
%information also needs to be added to make it more pragmatic.

%%% Local Variables:
%%% mode: latex
%%% TeX-master: "main"
%%% End:

%-------------------------------------------------------------------------------
\Urlmuskip=0mu plus 1mu
% bibliography section
\small\bibliography{main}
\bibliographystyle{unsrt}
% Appendix commented for now
%\input{appendix}
%%%%%%%%%%%%%%%%%%%%%%%%%%%%%%%%%%%%%%%%%%%%%%%%%%%%%%%%%%%%%%%%%%%%%%%%%%%%%%%%
\end{document}